\documentclass[a4paper,fleqn,usenatbib,useAMS,letters]{mnras}
\usepackage{newtxtext,newtxmath}
\usepackage[T1]{fontenc}
\usepackage{ae,aecompl}
\usepackage{graphicx}    
\usepackage{amsmath}     
\usepackage{multicol}    
\usepackage{bm}          
\usepackage{pdflscape}   
\usepackage{xspace}
\newcommand{\Msun}{\,$M_{\sun}$\xspace}

\newcommand{\kms}{\,km\,s$^{-1}$\xspace}
\newcommand{\ergs}{\,erg\,s$^{-1}$\xspace}

\newcommand{\HeI}{\ion{He}{i}\xspace}
\newcommand{\OI}{\ion{O}{i}\xspace}

\title[Luminous type Ic SN~2019stc] 
   {Origin of postmaximum bump in luminous type Ic SN~2019stc}
\author[N. N. Chugai \& V. P. Utrobin]{%
Nikolai N. Chugai$^{1}$\thanks{E-mail: nchugai@inasan.ru}
and
Victor P. Utrobin$^{1,2,3}$\\
$^{1}$Institute of Astronomy, Russian Academy of Sciences, Pyatnitskaya
      St. 48, 119017 Moscow, Russia\\
$^{2}$NRC `Kurchatov Institute' --
      Institute for Theoretical and Experimental Physics,
      B.~Cheremushkinskaya St. 25, 117218 Moscow, Russia \\
$^{3}$Max-Planck-Institut f\"ur Astrophysik, Karl-Schwarzschild-Str. 1,
      85748 Garching, Germany \\
}

\date{Accepted XXX. Received YYY; in original form ZZZ}

\pubyear{2021}

\begin{document}
\label{firstpage}
\pagerange{\pageref{firstpage}--\pageref{lastpage}}
\maketitle
%
\begin{abstract}
 We address the issue of the postmaximum bump observed in the light curve 
 of some superluminous supernovae.
We rule out the popular mechanism of a circumstellar interaction 
  suggested for the bump explanation.
Instead we propose that the postmaximum bump is caused by  
  the magnetar dipole field enhancement several months after the explosion.
The modeling of SN~2019stc light curve based on the thin shell approximation implies 
  that at the age of $\sim 90$ days the initial dipole magnetic field should be
   amplified by a factor of 2.8 to account for the postmaximum bump.
The specific mechanism for the field amplification of the newborn magnetar 
   on the timescale of several months has yet to be identified.
\end{abstract}

\section{Introduction} 
\label{sec:intro}
In the last decade, a new category of supernovae (SNe) has been discovered ---
   hydrogen-free superluminous supernovae (SLSNe-I) with the luminosity
   $>$3$\times10^{43}$\ergs \citep{Gal-Yam_2019}.
Their radiation is presumably powered by a magnetar \citep{Maeda_2007,
   Kasen_2010, Woosley_2010, Kasen_2016}, possibly with some contribution of
   radioactive $^{56}$Ni \citep[e.g.][]{Gomez_2021}.
While a standard magnetar mechanism is able to describe the light curve,
   including the early brief ($<10$\,d) bump \citep{Kasen_2016}, it cannot
   account for the postmaximum bump seen in some SLSNe-I $1-2$ months after
   the main light maximum, viz, in SN~2019stc \citep{Gomez_2021}.
The occurrence of the postmaximum bump among SLSNe and its origin have been
   recently analyzed by \citet{Yan_2017} and \citet{Hosseinzadeh_2021} with
   the conclusion that the circumstellar (CS) interaction or the variability
   of a central engine could be responsible for the bump.
At the moment \citet{Hosseinzadeh_2021} do not find any evidence to favor one
   mechanism over another.
  
As a matter of fact, there is a strong argument against the CS interaction.
Indeed, the light curve of SN~2019hge (SLSN-I) demonstrates a clear-cut
   postmaximum bump similar to that of SN~2019stc, yet SN~2019hge spectra
   \citep{Yan_2020} show pronounced \HeI absorption lines through the extended
   period of time, including the epoch of the postmaximum bump.
The presence of \HeI absorptions at the bump stage means that the bulk of
   luminosity is generated in the ejecta interior, {\em and not by the CS
   interaction.}
This implies that the likely reason for the bump origin of SLSNe-I is
   a variability of a central engine.

Here we propose and explore an alternative conjecture that the postmaximum bump
   is related to the enhancement of the dipole magnetic field of the magnetar
   that causes the higher magnetar luminosity responsible for the bump.
At the moment, we are not able to identify a specific mechanism for the field
   enhancement.
Yet the conjecture finds an indirect support in the variety of mechanisms for
   the magnetic field amplification of a newborn neutron star by means of, e.g.,
   convective dynamo \citep{Raynaud_2020},
   shear-Hall instability \citep{Kondic_2011},
   precession-driven amplification \citep{Lander_2021}, and
   reconfiguration of a partially submerged magnetic field \citep{Torres_2016}.
In the absence of reasonable alternative the effect of the enhancement of
   the dipole magnetic field should be taken as a viable possibility and
   must be explored.
We will apply our conjecture to the description of the light curve of the recent
   well-observed luminous type Ic supernova  SN~2019stc \citep{Gomez_2021}
   that shows a conspicuous postmaximum bump.

\section{Scenario and model overview}
\label{sec:model}
We suggest that the explosion of a WR star at $\sim$50\,d before the light
   maximum ($\sim$10\,d before the SN discovery) with the energy
   $E\sim10^{51}$\,erg ejects a freely expanding envelope with the mass
   $M\sim10$\Msun and the density distribution $\rho = \rho_0(t)/(1 + u^9)$,
   where $u = v/v_0$, while $\rho_0$ and $v_0$ are determined by $E$ and $M$.   
A rapidly rotating newborn neutron star loses the rotational energy presumably 
   via the magnetized relativistic wind \citep{Kennel_1984} that inflates
   the bubble with the energy $E_b$, the volume $V_b$, and the relativistic
   pressure $p \sim (1/3)E_b/V_b$.
The bubble expansion sweeps up the gas of the freely expanding envelope into
   a thin shell.
This shell however is liable to the Rayleigh-Taylor (RT) instability that is
   absent in our model.
Yet we take into account the effect of the RT fragmentation of the swept-up shell
  in the computation of the optical depth assuming homogeneous distribution of
  the swept-up mass in the bubble.
The internal energy of the bubble is spent on the pressure work, and on the
   escaping radiation responsible for the observed bolometric luminosity.

The issue of the conversion of a relativistic wind, dominated by Poynting flux,
   into the wind with the significant fraction of the particle thermal energy
   \citep[e.g.][]{Coroniti_2017} that powers the light curve is beyond our task.
Below we simply admit that some fraction $\eta$ of the magnetar luminosity $L_m$
   goes into the thermal energy in the form of radiation, while the remaining
   fraction $(1 - \eta)$ resides in the magnetic energy.
We also admit that the value $\eta$ can be variable \citep{Kasen_2016}.  
Both radiation and magnetic energy compose the bubble internal energy 
   $E_b = E_r + E_m$ with the relativistic pressure $p$ assumed to be uniform
   in the spherical volume $V_b$. 

The magnetar luminosity is determined by the magnetic dipole radiation of
   the rotating dipole with the angular frequency $\Omega$:
   $L_m = (2/3)(\mu \sin{\chi})^2\Omega^4/c^3 = (2/3)B_0^2R_{ns}^6\Omega^4/c^3$,
   where $\chi$ is the angle between the magnetic moment $\mu$ and the rotational
   axis and $B_0$ is the equatorial component of the surface magnetic field
   \citep{Landau_1975}.
More relevant is the expression for the spin-down losses obtained via
   magnetohydrodynamic simulations $L_m = \mu^2\Omega^4(1 + \sin^2{\chi})^2)/c^3$
   \citep{Philippov_2015}. 
Yet we employ the expression for the magnetic dipole radiation that is widely
   used for the magnetar in SLSNe-I \citep[e.g.][]{Chen_2020}.
This means that the magnetic field estimate can differ from a realistic value
   by a factor of the order of unity.
   
The equations of motion, magnetic and radiation energy, and mass conservation
   in the thin shell approximation with the shell mass $M_s$, the radius $r_s$,
   the velocity $v_s$, and the undisturbed SN density $\rho(t,r)$ at
   the radius $r_s$ read
\begin{eqnarray}
M_s\frac{dv_s}{dt} &=& 4\pi r_s^2[p - \rho(v_s - \frac{r_s}{t})^2] \\
\frac{dE_m}{dt} &=& -\frac{E_m v_s}{r_s} + (1- \eta)L_m \\
\frac{dE_r}{dt} &=& -\frac{E_r v_s}{r_s} - L_b + \eta L_m \\  
\frac{dM_s}{dt} &=& 4\pi r_s^2\rho(v_s - \frac{r_s}{t}) 
\end{eqnarray}  
The luminosity of the radiation escaping from the bubble is $L_b = E_r/t_d$,
   where $t_d$ is the diffusion time for the bubble photons taken to be equal
   to the average escape time for the case of a central source in the homogeneous
   envelope of the radius $r$ and optical depth $\tau$, i.e.,
   $t_d = t_{esc} = (r/c)\tau/2$ \citep[e.g.][]{ST_1980}. 
In our case $r = \mbox{max}(v_0t, r_s)$ and $\tau$ is the total  
  optical depth
   of the shell plus external ejecta $\tau = \tau_s + \tau_e$.

The supernova luminosity includes also the radiation produced by the radiative
   shock driven by the swept up shell.
In this respect our model is similar to that of \cite{Kasen_2016} and differs
   from the model by \cite{Gomez_2021} that does not include this luminosity
   component.
The shock luminosity is approximated as
\begin{equation}
L_s = 2\pi r_s^2\rho\left(v_s-\frac{r_s}{t}\right)^3\frac{t}{(t + t_e)}\,,
\end{equation}
   where $t_e$ is the diffusion time for the external ejecta
   $t_e = \mbox{max}(v_0t,r_s)\tau_e/2$.
The total supernova luminosity is thus $L_{sn} = L_b + L_s$.
We include the radioactive $^{56}$Ni presumably residing in the deep interior
   of the ejecta. 
We use the constant opacity $k = 0.1$\,cm$^2$\,g$^{-1}$ that is twice as lower
   compared to the value adopted earlier \citep{Gomez_2021}.

The additional component of the dipole field responsible for the 
  postmaximum bump can be described via a smooth step function $\theta(t)$ 
  and a relative amplitude $h$ with respect to the initial field $B_0$ as $B_1(t) = hB_0\theta(t)$. 
The function $\theta(t)$ is specified by the moment of the additional field turn on $t_1$ 
  and the field rise time $t_2$. 
We find it convenient to set $\theta(t)$ as
\begin{equation}
\theta(x)=\left\{
\begin{array}{lr}
0             & \mbox{if $x<0$}\,\phantom{,} \\
 x^4/(1+x^4) & \mbox{if $x\ge 0$}\,, \\
\end{array} \right.
\end{equation}
   where $x = (t-t_1)/t_2$.
The overall dipole field is thus $B = B_0[1 + h\theta(x)]$.
Values of $t_1$, $t_2$, and $h$ are recovered from the optimal fit of the light curve.
\begin{table}
\centering 
\caption[]{Model parameters}
\label{tab:par}
\begin{tabular}{p{1.cm}|p{1.2cm}|p{1.2cm}|p{1.2cm}}
\hline
Parameter & Units          & mag8  & mag8ni \\
\hline
 $E$          & $10^{51}$\,erg & 0.8  & 0.8 \\
 $M$          & $M_{\odot}$    & 8    & 8 \\
 $B_0$        & $10^{13}$\,G   & 5    & 5 \\
 $p_0$        & $10^{-3}$\,s   & 2.5  & 2.5 \\
 $\eta$       &                & 0.46 & 0.43 \\
 $h$          &                & 1.8  & 1.8 \\
 $M_{\mathrm{Ni}}$     & $M_{\sun}$      & 0    & 0.2 \\
 $t_1$        & days           & 92   & 92  \\
 $t_2$        & days           & 25   & 25  \\
 $t_m$        & days           & 52.5 & 52  \\ 
\hline
 $E_f^{\dag}$ & $10^{51}$\,erg & 3.9  & 3.9 \\
\hline
\parbox[]{6cm}{\small $^{\dag}$ Final kinetic energy. }
\end{tabular}
\end{table}
\begin{figure}
   \includegraphics[width=\columnwidth, clip, trim=40 110 0 0]{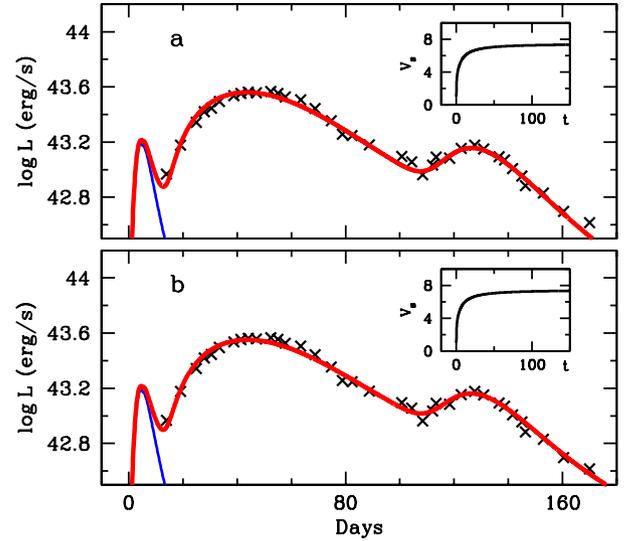}
   \caption{
   Model bolometric light curve of SN~2019stc (\emph{red\/}) overplotted on
      the observational bolometric light curve (\emph{crosses\/})
      reported by \citet{Gomez_2021}. 
   Panels (a) and (b) correspond to models mag8 and mag8ni, respectively
      (Table~\ref{tab:par}). 
   Thin \emph{blue} line shows the contribution of the luminosity of the
      radiative shock driven by the expanding bubble.
   Inset demonstrates the evolution of the thin shell speed in units of 1000\kms.
   }
   \label{fig:blcrv}
\end{figure}
%

\section{Light curve model}
\label{sec:lcurve}
The bolometric light curve of SN~2019stc analyzed here is recovered by
   \cite{Gomez_2021}.
We present two models: model mag8 without radioactive $^{56}$Ni and model mag8ni
   with a maximal amount of $^{56}$Ni of 0.2\Msun (Table~\ref{tab:par})
   consistent with the observational light curve (Fig.~\ref{fig:blcrv}).
The adopted neutron star mass, its radius and moment of inertia are
   $M_{ns} = 1.4$\Msun, $R_{ns}=12$\,km, and $I = 1.2\times10^{45}$\,g\,cm$^2$,
   respectively.
Table \ref{tab:par} contains from top to bottom: the explosion energy, the ejecta
   mass, the dipole surface magnetic field, the initial rotation period,
   the thermalized fraction of the relativistic wind, the relative amplitude of
   the additional component of the dipole field, the radioactive $^{56}$Ni mass,
   the turn on moment of the field amplification, the rise time of the field,
   and the time of the light maximum after the explosion.
Shown at the bottom is the final kinetic energy of the ejecta that exceeds the
   explosion energy by the $pdV$ work produced by the magnetar driven bubble.  
The relative error of parameters of the additional field component 
	$t_1$, $t_2$, and $h$ is about 5\%.  

Both models fit the observational light curve fairly well and provide required
   velocity (7000\kms) of the bubble boundary on +15\,d after the light maximum.
The initial peak of the model light curve is related to the radiative shock
   driven by the bubble expansion that is absent in the model of
   \cite{Gomez_2021}.
This shock is identified by \citet{Kasen_2016} with the early bump seen in
   several SLSNe-I.
The early bump generally affects the rise of the light towards the main maximum.
To describe the light curve of SN~2019stc, one needs to adopt that during the
   first 12 days the magnetar wind has the low thermalization parameter is low
   $\eta = 0.053$ with the subsequent transition to the larger value $\eta = 0.46$ and 
   $\eta = 0.43$ in models mag8 and mag8ni, respectively.
   
Disregarding these details we stress our major result: the postmaximum bump
   is well reproduced in the model with the initial dipole field of the magnetar
   $5\times10^{13}$\,G that is enhanced to $1.4\times10^{14}$\,G at about
   100\,d after the explosion.
We believe that the proposed scenario for the origin of the postmaximum bump is 
  applicable to the most of similar cases of SLSNe-I, including SN~2019hge that 
  permits us to abandon the alternative mechanism based on the circumstellar interaction.

Our simple model has an apparent drawback, viz., at the postmaximum stage it predicts
   that expanding ejecta should contain ``empty'' bubble with the velocity of
   $\sim$7000\kms. 
Formally this suggests that the profile of an emission line, e.g., \OI 7774\,\AA\
    should be flat-topped in the range of $\sim \pm7000$\kms, which is not
    the case \citep[cf.][]{Gomez_2021}.
In reality, the swept-up shell accelerated by the relativistic bubble is liable
   to the RT instability \citep{CheFra_1992}.
This results in a picture of light bubbles penetrating the 
  ejecta along with spikes of a dense shell material trailed behind
   \citep[e.g.][]{Layzer_1955}.
The spikes are subject to the stripping due to a shear flow, so      
 the expanding relativistic bubble turns out filled by the ejecta material as demonstrates by 3D-simulation  \citep{Chen_2020}. 
These 3D-hydrodynamic effects thus should resolve the issue of the bubble 
 emptiness of our one-dimensional model.
\section{Conclusion}
\label{sec:concl}
We found strong argument against the CS interaction as a mechanism for the
   bump emergence, which means that the only viable possibility is the
   variability of the central engine.
We explored the proposed conjecture that the postmaximum bump of the SN~2019stc
   bolometric light curve is related to the magnetar dipole field increase at
   about three months after the explosion.
The modeling suggests that the postmaximum bump is reproduced, if the dipole
   magnetic field is increased by a factor of $\sim$2.8 on the time scale
   of $\sim$25\,d starting from day 92 after the explosion. 
   
The physics behind the significant field amplification in three months after the core collapse  has yet to be identified. 
Presumably it could be related to mechanisms mentioned in the Introduction: either the convective   
	   field amplification or the reconfiguration of a partially submerged magnetic field.   

\section*{Acknowledgements}
We thank Maxim Barkov for discussions on a newborn magnetar field transformation.
The reported study was funded by RFBR and DFG, project number 21-52-12032.

\section*{Data Availability}
The data underlying this article will be shared on reasonable request to
   the corresponding author.

\bibliographystyle{mnras}
\bibliography{SN19stc_refs} 
\bsp	
\label{lastpage}
\end{document}